\documentclass[runningheads]{svmult}

\usepackage{makeidx}   
\usepackage{graphicx}  
\usepackage{subeqnar}  
\usepackage{multicol}  
\usepackage{physprbb}  
\makeindex             

\def\beq{\begin{equation}}
\def\eeq{\end{equation}}
\def\bea{\begin{eqnarray}}
\def\eea{\end{eqnarray}}
\def\bq{\begin{quote}}

\def\ga{\gamma}

\def\De{\Delta}

\def\prt{\partial}

\def\half{{\textstyle{1\over 2}}}
\def\frac#1#2{{\textstyle{{#1}\over {#2}}}}

\def\lsim{\mathrel{\rlap{\lower4pt\hbox{\hskip1pt$\sim$}}
    \raise1pt\hbox{$<$}}}
\def\gsim{\mathrel{\rlap{\lower4pt\hbox{\hskip1pt$\sim$}}
    \raise1pt\hbox{$>$}}}
\def\sqr#1#2{{\vcenter{\vbox{\hrule height.#2pt
         \hbox{\vrule width.#2pt height#1pt \kern#1pt
         \vrule width.#2pt}
         \hrule height.#2pt}}}}

\def\rl{\stackrel{\leftrightarrow}{\hskip1pt\prt^{\nu}}}

\def\ASAS{{\it Astron. and Astrophys.} }

\def\gappeq{\mathrel{\rlap {\raise.5ex\hbox{$>$}}
{\lower.5ex\hbox{$\sim$}}}}
 
\def\lappeq{\mathrel{\rlap{\raise.5ex\hbox{$<$}}
{\lower.5ex\hbox{$\sim$}}}}
 
\begin{document}
\title*{Threshold Effects and Lorentz Symmetry}
\toctitle{Threshold Effects and Lorentz Symmetry}
%
%
\titlerunning{Threshold Effects and Lorentz Symmetry}
%
\author{Orfeu Bertolami}
\authorrunning{Orfeu Bertolami}
%
%
\institute{Instituto Superior T\'{e}cnico, Depto. F\'\i sica,\\
           Av. Rovisco Pais, 1049-001 Lisboa, Portugal}

\maketitle              

\begin{abstract}
Evidence on the violation of Lorentz symmetry arises from the observation of cosmic rays 
with energies beyond the GZK cutoff, $E_{GZK} \simeq 4 \times 10^{19}~eV$, from the apparent 
transparency of the Universe to the propagation of high energy gamma radiation and from 
the stability of pions in air showers. These three paradoxes can be
explained through deformations of the relativistic dispersion relation. 
Theoretical ideas aimed to understand how Lorentz symmetry may be broken 
and phenomenologically interesting deformations of the relativistic dispersion relation 
may arise are briefly discussed.  
\end{abstract}

\section{Introduction}
Invariance under Lorentz transformations is one of the most fundamental 
symmetries of physics and is a key feature of 
all known physical theories. However, recently, evidence has emerged that this symmetry may not be 
respected in at least three different phenomena:

\vskip 0.3cm

\noindent
i) Observation of ultra-high energy cosmic rays with energies \cite{Hayashida,Bird,Lawrence,Efimov} 
beyond the Greisen-Zatsepin-Kuzmin (GZK) 
cutoff, $E_{GZK} \simeq 4 \times 10^{19}~eV$ \cite{Greisen}. These events, besides challenging our   
knowledge of mechanisms that allow accelerating cosmic 
particles to such high energies, may imply in a violation of Lorentz invariance as 
only through this 
violation that new threshold effects may arise and 
the resonant scattering reactions with photons of the Cosmic Microwave 
Background (CMB), e.g. 
$p + \ga_{2.73K} \to \De_{1232}$, are suppressed 
\cite{Sato1,Coleman,Mestres,Bertolami1}. An astrophysical solution to this paradox is 
possible and imply identifying viable sources at distances within, $D_{Source} \lsim 
50 - 100~Mpc$ \cite{Stecker,Hill}, so that the travelling time of the emitted particles 
is shorter than the attenuation time due to particle photoproduction on the CMB. 
Given the energy of the observed ultra-high energy cosmic rays (UHECRs) and the Hilla's criteria 
\cite{Hillas} on the 
energy, size and intensity of the magnetic field to accelerate protons, 
$E_{18} \le {1 \over 2} \beta B(\mu G) L(kpc)$ - where $E_{18}$ is 
the maximum energy measured in units of $10^{18}~eV$, $\beta$ the velocity 
of the shock wave relative to $c$ - it
implies that, within a volume of radius 
$50 - 100~Mpc$ about the Earth, only neutron stars, active galactic 
nuclei, gamma-ray bursts and cluster of galaxies 
are feasible acceleration sites \cite{Hillas,Cronin}. 
This type of solution has been recently suggested 
in Ref. \cite{Farrar}, where it was also argued that the near isotropy 
of the arrival directions of the observed UHECRs can be attributed to 
extragalactical magnetic fields near the Milky Way that are strong enough to 
deflect and isotropize the incoming directions of UHECRs from sources within 
$D_{Source}$. This is a debatable issue and it is worth bearing in mind 
that scenarios for the origin of UHECRs tend to generate anisotropies (see e.g. Ref. 
\cite{Olinto}). A further objection against this proposal 
is the mismatch in the energy fluxes of observed UHECRs 
and of the potential sources as well as the lack of spatial correlations 
between observed UHECRs and candidate sources (see e.g. \cite{Bertolami2} and references therein).

\vskip 0.3cm

\noindent
ii) Observations of gamma radiation with energies beyond
$20~TeV$ from distant sources such as Markarian 421 and Markarian 501 blazars
\cite{Krennrich,Aharonian,Nikishov}. These observations 
suggest a violation of Lorentz invariance as otherwise, 
due to pair creation, there should exist a strong attenuation of fluxes beyond $100~Mpc$ 
of $\gamma$-rays with energies higher 
than $10~TeV$ by the diffuse extragalactic background of infrared photons \cite{Amelino1,Coppi,Kifune,Kluzniak}.

\vskip 0.3cm

\noindent
iii) Studies of longitudinal 
evolution of air showers produced by ultra high-energy hadronic particles seem to 
suggest that pions are more stable than expected \cite{Antonov}. 

\vskip 0.3cm

\noindent
Violations of the Lorentz symmetry may lead to other threshold effects 
associated to asymmetric momenta in pair creation, photon stability, alternative  
\v{C}erenkov effects, etc \cite{Konopka,Jacobsen}.
 
On the theoretical front, work in the context of string/M-theory has shown 
that Lorentz symmetry can be spontaneously broken due to
non-trivial solutions in string field theory  \cite{Kostelecky1}, 
from interactions that may arise in braneworld scenarios where our $3$-brane is dynamical 
\cite{Dvali}, in loop quantum gravity \cite{Gambini,Alfaro}, in noncommutative field theories 
\footnote{We mention however, that  in a model where a scalar field is coupled to gravity, Lorentz 
invariance may still hold, at least at first non-trivial order in perturbation theory of the 
noncommutative parameter \cite{Bertolami5}.} 
\cite{Carroll}, and in quantum gravity inspired spacetime foam 
scenarios \cite{Garay}. 
The resulting novel
interactions may have striking implications at low-energy 
\cite{Bertolami6,Bertolami3,Amelino2,Mavromatos,Aloisio}.
Putative violations of the Lorentz invariance may also lead to
the breaking of CPT symmetry \cite{Kostelecky2}. 
An extension of the Standard Model (SM) that incorporates violations 
of Lorentz and CPT 
symmetries was developed in Ref. \cite{Colladay2}.

\section{Possible Solutios for the Observational Paradoxes and Experimental Bounds}

Potential violations of fundamental symmetries naturally raise the question of 
how to experimentally verify them. In the case of CPT symmetry, its violation can
be experimentally tested by various methods, such as for instance, via
neutral-meson experiments \cite{Colladay1}, Penning-trap 
measurements and hydrogen-antihydrogen spectroscopy \cite{Bluhm}. 
The breaking of CPT symmetry also allows for 
a mechanism to generate the baryon asymmetry of the Universe \cite{Bertolami4}.
In what concerns Lorentz symmetry, astrophysics plays, as we have already seen, an 
essential role. Moreover, it will soon 
be possible to make correlated
astrophysical observations involving high-energy radiation and, for instance, 
neutrinos, which will make viable direct astrophysical tests of 
Lorentz invariance \cite{Bertolami1,Amelino1,Biller} .

The tighest experimental limit on the extent of which Lorentz invariance is an 
exact symmetry arises from measurements of 
the time dependence of the quadrupole splitting of nuclear Zeeman levels
along Earth's orbit. Experiments of this nature can yield an impressive 
upper limit on deviations from the Lorentz invariance, 
$\delta < 3 \times 10^{-22}$ \cite{Lamoreaux}, 
and even more stringent bounds according to Ref.  
\cite{Kostelecky3}.

On very broad terms, proposals to explain the three abovementioned paradoxes rely on  
deformations of the relativistic dispersion relation, that can be written, for a particle species  
$a$, as:

\begin{equation}
E_a^2 = p_a^2 c_a^2 + m_a^2 c_a^4 + F(E_a, p_a, m_a, c_a)
\quad,
\label{1}
\end{equation}    
where $c_a$ is the maximal attainable velocity for particle $a$ and $F$ is a function of $c_a$ and of the relevant 
kinematical variables . 

For instance, Coleman and Glashow \cite{Coleman} proposed to explain the observation of cosmic rays beyond 
the GZK limit assuming that each particle has its own maximal attainable velocity and a vanishing function F. 
This is achieved studying the relevant interaction between a CMB photon and 
a proton primary yielding the $\Delta(1224)$ hadronic resonance. A tiny difference between the maximal 
attainable velocities, $c_p - c_{\Delta} \equiv \epsilon_{p \Delta} 
\simeq 1.7 \times 10^{-25}~c$, can explain the events beyond the GZK 
cutoff. This bound is three orders of magnitude more stringent than the experimental one. 
A bound from the search of neutrino oscillations can also be found, even though less stringent, 
$\vert \epsilon \vert \lsim few \times 10^{-22}~c$ \cite{Brucker}. Interestingly 
these limits can be turned into bounds on parameters of the 
Lorentz-violating extension of the SM \cite{Bertolami1}. 
As discussed in Ref. \cite{Bertolami1}, a characteristic feature of the Lorentz violating 
extension of the SM of Ref. \cite{Colladay2} is that it gives origin  
to a time delay, $\Delta t$, in the arrival of signals 
brought by different particles that is energy independent, in 
opposition to what is expected from other models (see \cite{Amelino2} and 
references therein), and 
has a dependence on the chirality of the particles involved as well:

\begin{equation}
\Delta t \simeq {D \over c} [(c_{00} \pm d_{00})_{i} - 
(c_{00} \pm d_{00})_{j}] 
\quad,
\label{2}
\end{equation}
where $c_{00}$ and $d_{00}$ are the time-like components of the CPT-even 
flavour-dependent parameters that have to be added the fermion sector of the 
SM so to exhbit Lorentz-violating interactions \cite{Colladay2}

\begin{equation}
{\cal L}^{\rm CPT-even}_{\rm Fermion} = \half i c_{\mu\nu} \overline{\psi}
\gamma^{\mu} \rl \psi + \half i d_{\mu\nu} \overline{\psi} \gamma_5
\gamma^{\mu} \rl  \psi 
\quad.
\label{3}
\end{equation}
The $\pm$ signs in Eq. (\ref{2}) arise from the fact that parameter 
$ d_{\mu\nu}$ depends on the chirality of the particles in question, and $D$ is 
the proper distance of the source.

The function $F$ arising from this SM Lorentz violating extension is given by 
\cite{Bertolami1}:

\begin{equation}
F = -2 c_{00}E^2 \pm 2 d_{00}Ep
\quad, 
\label{4}
\end{equation}
with $c_a = c$ for all particles.

It has been argued that some quantum gravity and  
stringy inspired models (see \cite{Alfaro,Amelino2,Mavromatos,Aloisio}) lead to modifications of 
the dispersion relation of the following form:

\begin{equation}
F = - k_a {p_a^3 \over M_P}
\quad, 
\label{5}
\end{equation}
where $k_a$ is a constant and $M_P$ is Planck's mass. This deformation can explain the three 
discussed paradoxes \cite{Alfaro,Amelino2,Mavromatos,Aloisio,Konopka}.

At very high energies, deformation (\ref{5}) can be approximately written as

\begin{equation}
F \simeq - {E^3 \over E_{QG}}
\quad, 
\label{6}
\end{equation}
where $E_{QG}$ is a quantum gravity scale. For photons this leads to the following dispersion relation 

\begin{equation}
pc = E \sqrt{1 + {E \over E_{QG}}}
\quad, 
\label{7}
\end{equation}
from which bounds on the quantum gravity scale 
\cite{Amelino1,Amelino2,Mavromatos,Biller} can be astrophysically determined,
the most stringent being \cite{Amelino3}

\begin{equation}
E_{QG} > 4 \times 10^{18}~GeV 
\quad. 
\label{8}
\end{equation}

Another aspect of the problem of violating a fundamental symmetry like Lorentz invariance 
concerns gravity.
A putative violation of Lorentz symmetry renews the interest in gravity theories 
that have intrinsically built in this feature. From the point of the  
post-Newtonian parametrization the theory that most closely 
resembles General Relativity is Rosen's bimetric 
theory \footnote{This theory has some difficulties as in its simplest form
it does not admit black hole solutions and it is unclear to which extent it is compatible with cosmology.} 
\cite{Rosen}. Indeed, this theory shares with General 
Relativity the same values for all post-Newtonian parameters \cite{Will}

\beq
\beta = \gamma = 1~~;~~\alpha_{1} = \alpha_{3} 
= \zeta_{1} =  \zeta_{2} = \zeta_{3} = \zeta_{4} = \xi = 0
\quad,
\label{9}
\eeq   
except for parameter $\alpha_{2}$ that signals the presence preferred-frame 
effects (Lorentz invariance violation) in the $g_{00}$ and $g_{0i}$ components of the metric. 
Naturally, this parameter vanishes in General Relativity, but in 
Rosen's bimetric theory it is given by

\beq
\alpha_{2} = {f_0 \over f_1} - 1
\quad,
\label{10}
\eeq   
where $f_0$ and $f_1$ are the asymptotic values of the components of the 
metric in the Universe rest frame, i.e. $g_{\mu \nu}^{(0)} = 
diag(- f_0, f_1, f_1, f_1)$, 
which must be close to the Minkowski metric. 
A non-vanishing $\alpha_{2}$ implies that angular 
momentum is not conserved. 
Bounds on this parameter are obtained from the resulting
anomalous torques on the Sun, whose absence reveals that 
$\alpha_{2} < 4 \times 10^{-7}$ \cite{Nordtvedt}. It is worth remarking that 
the other parameters leading to preferred-frame effects are bound by the pulsar 
PSR J2317+1439 data, that lead to $\alpha_{1} < 2 \times 10^{-4}$ 
\cite{Will}, and 
from the average on the pulse period of millisecond  
pulsars, which gives 
$\alpha_{3} < 2.2 \times 10^{-20}$ \cite{BellD}. It is clear that Rosen's theory 
deserves a closer examination. 

\section{Conclusions and Outlook}

Lorentz and CPT symmetries may be spontaneously broken in string theory 
and in some quantum gravity inspired models. Modifications to 
the relativistic dispersion relation arising from these models allow 
for explaning the three paradoxes associated to threshold effects in ultra high-energy 
cosmic rays, pair creation in the propagation of $TeV$ photons and its interaction with 
the diffuse gamma radiation background, and the longitudinal evolution of high energy hadronic 
particles in extensive air showers. Confirmation that these phenomena signal the breaking of
Lorentz symmetry is an exciting prospect as it would constitute in an inequivocal 
indication of physics beyond the SM. Near future observations that will be 
carried by extensive detectors such as by the Auger Observatory \cite{Auger} 
may unfold interesting questions and challenges to theory.


\begin{thebibliography}{8.}
\addcontentsline{toc}{section}{References}



\bibitem{Hayashida} N. Hayashida et al., (AGASA Collab.), Phys. Rev. Lett. \textbf{73},  3491 
(1994);

M. Takeda et al., (AGASA Collab.), Phys. Rev. Lett. \textbf{81},  1163 
(1998).



\bibitem{Bird}  D.J. Bird et al., (Fly's Eye Collab.),  Phys. Rev. Lett. \textbf{71}, 3401 
(1993); Astrophys. J. \textbf{424}, 491 (1994); \textbf{441}, 144 (1995).



\bibitem{Lawrence}  M.A. Lawrence, R.J.O. Reid and A.A. Watson 
(Haverah Park Collab.), Journ. Phys. \textbf{G17}, 733 (1991).



\bibitem{Efimov} N.N. Efimov et al., (Yakutsk Collab.), ICRR Symposium on
Astrophysical Aspects of the Most Energetic Cosmic Rays, eds. N. Nagano and
F. Takahara (World Scientific, 1991).



\bibitem{Greisen} K. Greisen, Phys. Rev. Lett. \textbf{16}, 748 (1966);

G.T. Zatsepin, V.A. Kuzmin, JETP Lett. \textbf{41}, 78 (1966).



\bibitem{Sato1} H. Sato, T. Tati, Prog. Theor. Phys. {\bf 47}, 1788 (1972).



\bibitem{Coleman} S. Coleman, S.L. Glashow, Phys. Lett. \textbf{B405}, 249 (1997); 
Phys. Rev. \textbf{D59}, 116008 (1999). 


\bibitem{Mestres} L. Gonzales - Mestres, hep-ph/9905430. 


\bibitem{Bertolami1} O. Bertolami, C.S. Carvalho, Phys. Rev. \textbf{D61}, 103002 
(2000).


\bibitem{Stecker} F.W. Stecker, Phys. Rev. Lett. \textbf{11}, 1016 (1968).


\bibitem{Hill} C.T. Hill, D. Schramm, T. Walker, Phys. Rev. \textbf{D36}, 1007 (1987).


\bibitem{Hillas} A.M. Hillas, Ann. Rev. Astron. Astrophys. \textbf{22}, 425 (1984). 


\bibitem{Cronin} J.W. Cronin, Rev. Mod. Phys. \textbf{71}, S165 (1999). 


\bibitem{Farrar} G.R. Farrar, T Piran, Phys. Rev. Lett. \textbf{84}, 3527 (2000).


\bibitem{Olinto} A.V. Olinto, astro-ph/0003013.


\bibitem{Bertolami2} O. Bertolami, Gen. Rel. Grav. \textbf{34}, 707 (2002).


\bibitem{Krennrich} F. Krennrich et al., Astrophys. J. \textbf{560}, L45 (2001).  


\bibitem{Aharonian} F.A. Aharonian et al., \ASAS \textbf{349}, 11A (1999).


\bibitem{Nikishov} A.N. Nikishov, Sov. Phys.  JETP \textbf{14}, 393 (1962);

J. Gould, G. Schreder, Phys. Rev. \textbf{\bf 155}, 1404 (1967);

F.W. Stecker, O.C. De Jager, M.H. Salmon, Astrophys. J. \textbf{\bf 390}, L49 (1992).


\bibitem{Amelino1} G. Amelino-Camelia, J. Ellis, N.E. Mavromatos, D.V. Nanopuolos, S. Sarkar, 
Nature \textbf{393}, 763 (1998).


\bibitem{Coppi} P.S. Coppi,  F.A. Aharonian, Astropart. Phys. \textbf{11}, 35 (1999).


\bibitem{Kifune} T. Kifune, Astrophys. J. Lett. \textbf{518}, L21 (1999).


\bibitem{Kluzniak} W. Kluzniak, astro-ph/9905308;

R.J. Protheroe, H. Meyer, Phys. Lett. \textbf{B493}, 1 (2000);

G. Amelino-Camelia, T. Piran, Phys. Rev. \textbf{D64}, 036005 (2001).



\bibitem{Antonov} E.E. Antonov et al., Pisma ZhETF \textbf{73}, 506 (2001).


\bibitem{Konopka} T.J. Konopka, S.A. Major, New J. Phys. \textbf{4}, 57 (2002).


\bibitem{Jacobsen} T. Jacobsen, S. Liberati, D. Mattingly, hep-ph/0209264. 


\bibitem{Kostelecky1} V.A. Kosteleck\'y, S. Samuel, 
Phys. Rev. \textbf{D39}, 683 (1989); Phys. Rev. Lett. \textbf{63}, 224 (1989).


\bibitem{Dvali} G. Dvali, M. Shifman, hep-th/9904021.


\bibitem{Gambini} R. Gambini, J. Pullin, Phys. Rev. \textbf{D59}, 124021 (1999).


\bibitem{Alfaro} J. Alfaro, H.A. Morales-Tecotl, L.F. Urrutia, Phys. Rev. Lett. \textbf{84}, 2183 (2000).


\bibitem{Carroll} S.M. Carroll, J.A. Harvey, V.A. Kosteleck\'y, C.D. Lane, T.Okamoto, 
Phys. Rev. Lett. \textbf{87}, 141601 (2001).


\bibitem{Bertolami5} O. Bertolami, L. Guisado, Phys. Rev. \textbf{D67}, 025001 (2003).


\bibitem{Garay} L.J. Garay, Phys. Rev. Lett. \textbf{80}, 2508 (1998).


\bibitem{Bertolami6}  O. Bertolami, Class. Quantum Grav. \textbf{14}, 2748 (1997).


\bibitem{Bertolami3} O. Bertolami, D.F. Mota, Phys. Lett. \textbf{B455}, 96 (1999). 


\bibitem{Sato2} H. Sato, astro-ph/0005218.


\bibitem{Amelino2} G. Amelino-Camelia, T. Piran, Phys. Lett. \textbf{B497}, 265 (2001).


\bibitem{Mavromatos} N. Mavromatos, gr-qc/0009045.


\bibitem{Aloisio} R. Aloisio, P. Blasi, P.L. Ghia, A.F. Grillo, astro-ph/0001258.


\bibitem{Kostelecky2} V.A. Kosteleck\'y, R. Potting, Phys. Rev. \textbf{D51}, 3923 
(1995); Phys. Lett. \textbf{B381}, 389 (1996).


\bibitem{Colladay2} D. Colladay, V.A. Kosteleck\'y, Phys. Rev.
\textbf{D55}, 6760 (1997); Phys. Rev. \textbf{D58}, 116002 (1998). 


\bibitem{Colladay1} D. Colladay, V.A. Kosteleck\'y,
Phys. Lett. \textbf{B344}, 259 (1995); Phys. Rev. \textbf{D52}, 6224 (1995);

V.A. Kosteleck\'y, R. Van Kooten, Phys. Rev. \textbf{D54}, 5585 (1996).


\bibitem{Bluhm} R. Bluhm, hep-ph/0006033.


\bibitem{Bertolami4} O. Bertolami, D. Colladay, V.A. Kosteleck\'y, R. Potting,
Phys. Lett. \textbf{B395}, 178 (1997).


\bibitem{Biller} S.D. Biller et al., Phys. Rev. Lett. \textbf{83}, 2108 (1999).


\bibitem{Lamoreaux} S.K. Lamoreaux, J.P. Jacobs, B.R. Heckel, F.J.  
Raab, E.N. Fortson, Phys. Rev. Lett. \textbf{57}, 3125 (1986).


\bibitem{Kostelecky3} V.A. Kosteleck\'y, C.D. Lane, Phys. Rev. \textbf{D60}, 116010 
(1999).


\bibitem{Brucker} E.B. Brucker et al., Phys. Rev. \textbf{D34}, 2183 (1986);

S.L. Glashow, A. Halprin, P.I. Krastev, C.N. Leung, J. Pantaleone, 
Phys. Rev. \textbf{D56}, 2433 (1997).


\bibitem{Amelino3} G. Amelino-Camelia, gr-qc/0212002.


\bibitem{Rosen} N. Rosen, Gen. Rel. Grav. \textbf{4}, 435 (1973); Ann. Phys. (New York) 
\textbf{84}, 455 (1974); Gen. Rel. Grav. \textbf{9}, 339 (1978).


\bibitem{Will} C.M. Will, ``Theory and Experiment in Gravitational  
Physics'' (Cambridge University Press, 1993); ``The Confrontation between 
General Relativity and Experiment: A 1998 Update, gr-qc/9811036.


\bibitem{Nordtvedt} K. Nordtvedt, Astrophys. J. \textbf{320}, 871 (1987).


\bibitem{BellD} J.F. Bell, T. Damour, Class. Quantum Grav.
\textbf{\bf 13}, 3121 (1996). 


\bibitem{Auger} L. Anchordoqui, T. Paul, S. Reucroft, J. Swain, astro-ph/0206072.






\end{thebibliography}
\end{document}